\documentclass[prb,preprint,amsmath,showpacs]{revtex4}
\usepackage{graphicx}

\begin{document}

\bibliographystyle{prsty}

\title {Anomalous electronic Raman scattering in $\bf Na_{x}CoO_2 \cdot y H_2O$}

\author {
 P. Lemmens$^{1,2}$,
 K.Y. Choi$^{3,4}$,
 V. Gnezdilov$^5$,
 E.Ya. Sherman$^6$
 D.P. Chen$^1$,
 C.T. Lin$^1$,
 F.C. Chou$^7$,
 B. Keimer$^1$}

\affiliation{
 $^1$ Max Planck Inst. for Solid State Research, D-70569 Stuttgart, Germany\\
 $^2$ Inst. for Physics of Condensed Matter, TU Braunschweig, D-38106
Braunschweig, Germany\\
 $^3$ 2. Physikalisches Inst., RWTH Aachen, D-52056 Aachen, Germany\\
  $^4$ Inst. for Materials Research, Tohoku University, Sendai 980-8577,
Japan\\
 $^5$ B. I. Verkin Inst. for Low Temperature Physics, NASU, 61164 Kharkov, Ukraine\\
 $^6$ Dept. of Physics, Univ. of Toronto, Toronto ON, M5S 1A7, Canada\\
 $^7$ Center for Materials Science and Engineering, MIT, Cambridge, MA 02139, USA
}\

\date{\today}

\begin{abstract}

Raman scattering experiments on $\rm Na_{x}CoO_2 \cdot yH_2O$ single crystals show a
broad electronic continuum with a pronounced peak around 100~$\rm cm^{-1}$ and a cutoff
at approximately 560~$\rm cm^{-1}$ over a wide range of doping levels. The electronic
Raman spectra in superconducting and non-superconducting samples are similar at room
temperature, but evolve in markedly different ways with decreasing temperature. For
superconducting samples, the low-energy spectral weight is depleted upon cooling below
$T^*$ $\rm \sim 150$ K, indicating a opening of a pseudogap that is not present in
non-superconducting materials. Weak additional phonon modes observed below $T^*$ suggest
that the pseudogap is associated with charge ordering.

\end{abstract}
 \maketitle



The hydrated cobaltate $\rm Na_xCoO_2\cdot yH_2O$ has recently been in the focus of
research on correlated electron systems, because it enables investigations of the
relationship between superconductivity (SC) \cite{takada03} and magnetic ordering on a
triangular lattice \cite{jorgensen03,lynn03}. Electronic correlations can be tuned
either by changing the Na content, leading to long-range antiferromagnetic order for
$\rm 0.75\leq x \leq 0.85$ ($\rm T_{N}=20$~K) \cite{motohashi03,bayrakci04} and a
metal-insulator transition at x=1/2 \cite{huang04,hwang04}, or by hydration leading to
superconductivity \cite{takada03,schaak03} in Na-poor samples with $\rm x \sim 1/3$ and
$\rm y \sim 4/3$.


For Na-rich cobaltates, angle-resolved photoemission experiments \cite{hasan04,yang04}
show a large hole-like, hexagonal Fermi surface dominated by Co $\rm t_{2g}$ states of
$\rm a_{g}$ symmetry \cite{singh00,koshibae03,johannes04}. Several anomalies are
observed in this doping range and for temperatures below approximately 150~K. Among them
are a break in a weakly dispersing quasiparticle band at an energy of $\sim$70~meV
$\equiv$ 560~$\rm cm^{-1}$, which was attributed to a bosonic mode of electronic or
lattice origin \cite{hasan04,yang04}; a T-linear resistivity; a colossal thermopower
\cite{wang03}; an anomalous linear frequency/energy dependency of the electronic
scattering rates \cite{caimi05}; and a polaronic mode \cite{bernhard04}. With increasing
temperature the low-energy quasiparticle peaks broaden substantially and become
incoherent \cite{hasan04,yang04}. Under such circumstances the nested Fermi surface with
a flat-band feature is expected to be susceptible to electronic instabilities, and
indeed charge ordering has been reported around x$\sim$0.5 \cite{ning04}. Furthermore,
SC states with unconventional/anisotropic order parameters may be expected
\cite{johannes04,tanaka03,baskaran03b,motrunich04}. It is therefore important to note
that in Na-poor, but still non-hydrated cobaltates (x$\approx$0.3), charge is more
delocalized and the Fermi liquid character of the quasiparticles appears to be recovered
with noticeable mass enhancement \cite{hasan04}.

However, very little spectroscopic information has thus far been reported on hydrated,
SC cobaltates. Here we report the observation of pronounced electronic Raman scattering
(RS) over a wide range of doping levels comprising the SC regime of the phase diagram.
This feature uncovers strong carrier scattering by bosonic modes, similar to that
observed in underdoped high temperature cuprates. The electronic RS continua in SC and
non-SC samples are similar at room temperature, but evolve in a markedly different way
as the temperature is lowered. Specifically, a pseudogap opens up below $\rm T \leq T^*
\sim 150$~K in SC, but not in non-SC samples. Weak additional phonon modes observed
below T$^*$ indicate that the pesudogap may be associated with charge ordering.

Single-crystal samples were prepared in an optical travelling-solvent floating-zone
furnace \cite{chen04,chou04}. This is to our knowledge the first systematic RS study of
such single crystals that are advantageous due to their easy cleavability and size
\cite{note}. The experiments ($\lambda$=514~nm, 6~mW, $\O$=100$\mu$m diameter laser
focus with parallel, in-plane light polarization) were performed in quasi-backscattering
geometry on the \emph{ab} surface of freshly cleaved single crystals. After cleavage,
the samples were instantaneously cooled down in He contact gas, in order to prevent
dehydration \cite{foo03} or Na ordering at elevated temperatures
\cite{huang04,zandbergen04}. We present RS data on four well characterized samples:
Three non-SC samples with $\rm y=0$ and $\rm x = 0.5$, 0.83, and $\leq$1, and one SC
sample with $\rm x = 0.35$ and $\rm y=1.3$.


\begin{figure}[t]
     \centering
     \includegraphics[height=8.5cm]{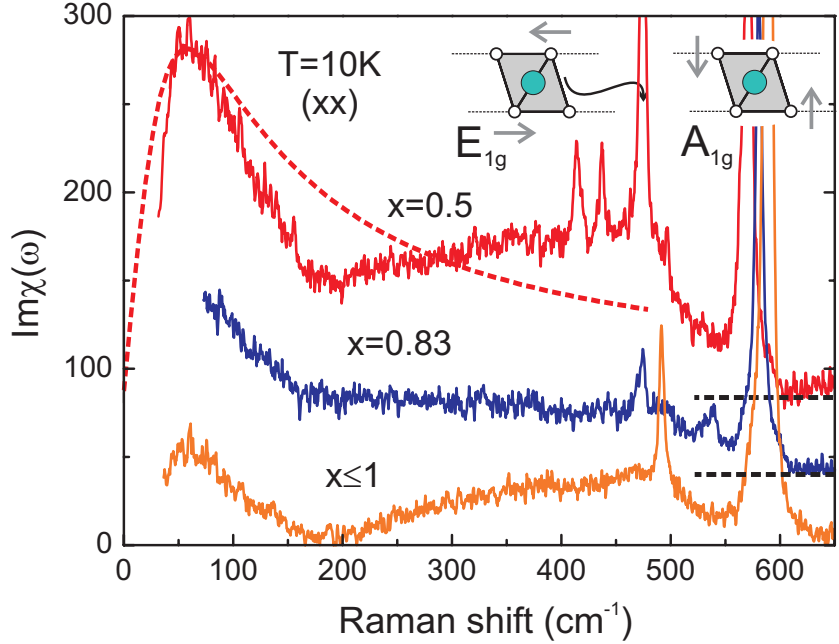}
          \caption{Raman cross section at T=10~K for x=0.5, 0.83,
1.0 with in-plane (xx) light polarization. The insets give oxygen displacements of the
main phonon modes using small (bigger) circles for oxygen (cobalt) ions. The curves are
shifted for clarity and the dashed lines give an estimate of the background scattering.
The curved dashed line corresponds to an electronic RS process with a single,
frequency-independent scattering rate, $\Gamma$=58~$\rm cm^{-1}$, that extrapolates
x=0.5 at low frequencies.
     }
     \label{elscatt}
 \end{figure}


Figure 1 shows the low temperature RS response $Im\chi$ (the measured RS cross section
corrected for the Bose thermal population factor) of three non-SC samples with Na
content $\rm x = 0.5$, 0.83, and $\sim 1$. All spectra show two strong phonon modes
around 580~$\rm cm^{-1}$ and 480~$\rm cm^{-1}$, which can be attributed to Raman-active
out-of-plane $\rm A_{1g}$ and in-plane $\rm E_{1g}$ vibrations, respectively, of oxygen
in the $\rm CoO_6$ octahedra \cite{iliev03,lemmens04b}. The eigenvectors are shown as
insets in Fig. 1. (Co sites do not contribute to Raman-active lattice vibrations due to
their inversion site symmetry.) The sharpness of the higher-frequency mode, whose
frequency softens systematically with decreasing x (inset in Fig. 2), testifies to the
homogeneity of the samples. The sample with $\rm x=0.5$, for which Na ordering and a
metal-insulator transition have been reported \cite{huang04,hwang04}, shows additional
modes at 413, 437, and 497~$\rm cm^{-1}$, close to the energy of the in-plane phonon.
This demonstrates the sensitivity of this phonon to structural and electronic ordering
processes. In contrast, the sample with $\rm x=0.83$ exhibits weak modes around 485 and
540~$\rm cm^{-1}$ that might be due to an admixture of inter-plane polarizations,
attributable to an $\rm E_{2g}$ mode involving both Na and O motions. Further, note that
no anomalous behavior of the phonon modes is observed in frequency and temperature
beyond lattice anharmonicities.

\begin{figure}[t]
\includegraphics[height=10cm]{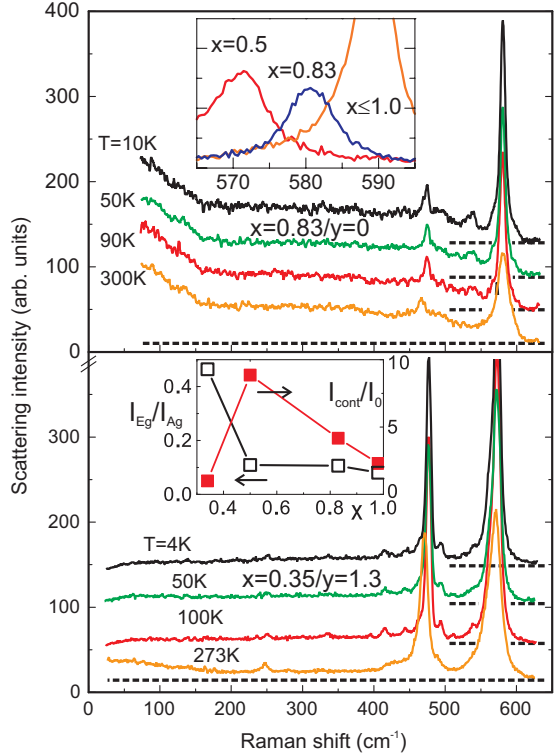}
\caption{RS intensity of $\rm Na_{0.83}CoO_2$ (upper panel) and $\rm Na_{0.34}CoO_2
\cdot 1.3D_2O$ (lower panel) in $xx$ light polarization with $x$ in the $ab$ plane of
the crystal. The upper inset shows a zoomed view on the out-of-plane $\rm A_{1g}$ mode.
The lower inset gives the intensity ratio of the in-plane to the out-of-plane phonon and
the integrated intensity of the continuum.  }
\end{figure}

The most striking aspect of the spectra shown in Fig. 1 is a continuum with a
pronounced, broad peak around $\sim 100~\rm cm^{-1}$. Above the peak, the intensity
levels off, and a plateau extends up to about 550~$\rm cm^{-1}$.  Since the crystals are
of high quality and show sharp phonon spectra, phonon density-of-states effects can be
ruled out as the origin of the continuum. The upper cutoff of the continuum remains at
roughly the same energy of 550~$\rm cm^{-1}$ irrespective of Na contents. This is in
excellent correspondence with the width of the quasiparticle band determined by
photoemission spectroscopy \cite{hasan04,yang04} as well as with the knee in the
reflectance seen by optical spectroscopy \cite{hwang04}. We hence attribute our
observations to electronic RS. Noticeably, the continuum intensity increases nearly
monotonically with increasing hole hoping, corresponding to decreasing Na content (lower
inset in Fig. 2). This effect corresponds to the increase of the Fermi surface with hole
doping that fulfills the Luttinger theorem \cite{yang04}. Deviation seen in the SC
sample will be discussed below.

Figure 2 shows in the upper panel the temperature evolution of the RS intensity of
antiferromagnetic $\rm Na_{0.83}CoO_2$ ($\mathrm{T_N}$=19.8~K). The data are
intentionally presented as raw, non-Bose corrected RS intensity. The weak temperature
dependence of the spectra demonstrates that the temperature dependence of the RS
scattering cross section approximately compensates the Bose thermal population factor.
The lower panel displays electronic Raman scattering in the superconducting sample of
composition $\rm Na_{0.34}CoO_2 \cdot 1.3D_2O$ and superconducting transition
temperature $\mathrm{T_{sc}}$=4.6~K. The overall shapes of the continua are similar at
room temperature.

A similar temperature dependence has been observed in RS on high temperature
superconductors. The nearly flat, frequency- and temperature-independent continuum has
been discussed in the framework of a marginal Fermi liquid state with a linear
dependence of the scattering rate on temperature and energy/frequency \cite{varma89}.
Thus, the unusual temperature dependence of the RS appears to be a consequence of strong
electronic correlations as in the high-T$_c$ cuprates. A linear temperature dependence
of the scattering rate has also been observed in IR and photoemission experiments on
cobaltates \cite{hasan04,caimi05}. However, there is some important difference. In the
cuprates both the electronic RS and the anomalous IR reflectance extend to 0.75 eV. In
contrast, the electronic RS of the cobaltates is strongly restricted to a low-frequency
region below $550~\rm cm^{-1}$($\approx 0.07$ eV). Notwithstanding the generally smaller
energy scale, this implies that the observed RS in the cobaltates strongly relies on a
mode having an energy scale of $600~\rm cm^{-1}$. Remarkably, photoemission and IR
spectroscopy measurements give evidence for a bosonic mode of electronic, lattice, or
magnetic origin in the same energy range \cite{hwang04,hasan04,yang04}. Although further
work is necessary to pin down its exact origin, it is clear that the interaction of
charge carriers with a bosonic mode puts constraints to their scattering frequency
range. This might be responsible for the low transition temperature in the cobaltates.

We will now focus on the sharply enhanced scattering seen below $\sim 150~\rm cm^{-1}$.
This energy scale corresponds to the temperature scale at which the quasiparticle
dynamics observed by photoemission becomes incoherent. At low frequencies, the lineshape
of the RS response of metals and semiconductors can often be approximated by a
Lorentzian as function of frequency $\omega$ and electronic scattering rate
$\Gamma(q,\omega)$: $\rm Im\chi(q,\omega) \propto (\omega \Gamma)/(\omega^{2}+
\Gamma^{2})$ \cite{zawadowski90}. This form provides a reasonable description of the
broad low-frequency peaks in the non-hydrated samples (Fig. 1) with a single, frequency
independent scattering rate $\Gamma \sim 60~\rm cm^{-1}$, not strongly dependent on
doping.

We note that similar observations have been made in ferromagnetic semiconductors, e.g.
in $\rm (Eu_{1-x},Gd_{x})O$, where the origin of the peak was identified as a magnetic
polaron excitation \cite{rho02}. Further, a similar peak was observed in the infrared
spectra of $\rm Na_{0.82} CoO_2$ (albeit at a somewhat larger frequency of $\sim 150 \rm
cm^{-1}$) and also interpreted as a magneto-polaron mode \cite{bernhard04}. We hence
tentatively ascribe the low-frequency peak in $\rm Na_{0.5} CoO_2$ to electronic RS on
polarons based on the $\rm t_{2g}$ states of Co. This may indicate that the scattering
rate $\Gamma$ in the samples with lower hole content originates from short-range charge
and spin fluctuations, rather than impurities as in conventional semiconductors. Recent
neutron scattering data have shown a marked broadening of the spin excitations in a
sample with $\rm x=0.82$ and $\rm y=0$, consistent with this idea \cite{bayrakci_SW}.

\begin{figure}[t]
     \centering
     \includegraphics[height=9.5cm]{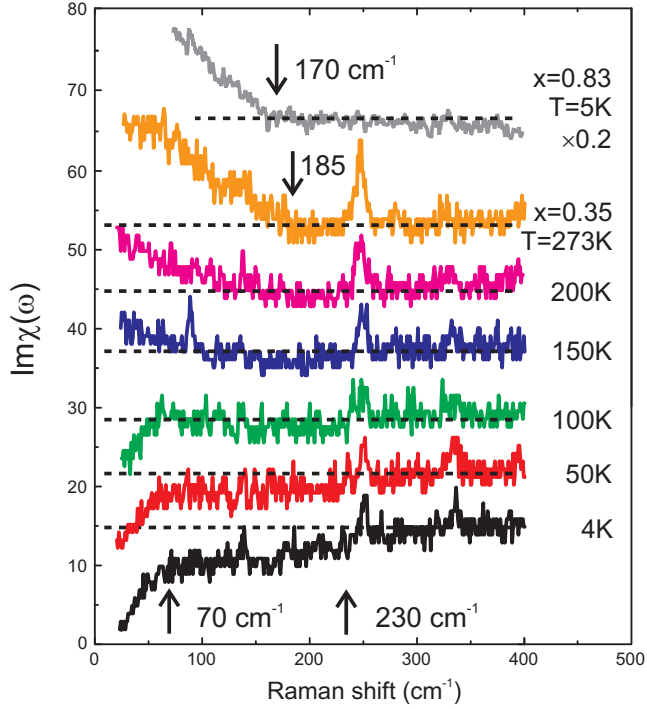}
\caption{Low energy RS cross section $Im\chi$ as function of
temperature for $\rm Na_{0.35}CoO_2 \cdot 1.3 D_2O$. The top curve
gives one data set for x=0.83 at T=4K scaled by a factor of 0.2.
The arrows mark typical energy scales.}
     \label{gap}
 \end{figure}

We next address the temperature dependence of the electronic Raman scattering of SC
samples. The overall intensity of the continuum (referenced to the intensities of the
dominant phonon modes, see lower inset in Fig. 2) is significantly weaker in the
hydrated, SC composition than in the non-SC samples. The strong intensity reduction
points to a reduced electronic scattering rate, which seems to be inconsistent with more
metallic behavior with higher hole doping. This should be ascribed to the fact that
Raman spectroscopy is sensitive to the low frequency carrier dynamics and scattering
processes without charge conservation. Thus, our Raman data can be interpreted in terms
of charge localizations, confirmed below.

The spectra of SC and non-SC samples show a very different evolution as the temperature
is lowered. This is highlighted in Fig. 3, where the low-energy Im$\chi$ is presented
for samples with x=0.35 and x=0.83. At high temperatures both spectra show similar kinks
at 185 and $170~\rm cm^{-1}$, respectively. While the kink position is temperature
independent in the non-SC sample (Fig. 2a), it shifts to lower energies with decreasing
temperature in the SC sample. For T$<$150~K, the low-energy spectral weight is sharply
suppressed, opening up a ``pseudogap" that is not present in samples with $\rm x \geq
0.5$. At 4 K, the spectral depletion is most pronounced at energies below $\sim 70~\rm
cm^{-1}$, but extends up to 230~$\mathrm{cm^{-1}}$, only a factor of two smaller than
the high-energy cutoff of the continuum. This implies that the pseudogap energy is not
only large compared to the SC transition temperature, but that it constitutes a
significant fraction of the energy scales characterizing the electronic structure. The
upper panel of Fig. 4 shows that the low-energy scattering rate $\rm \Gamma(T)$/$\rm
\Gamma(273K)$ determined from the inverse of the slope forms a maximum at $T^*$$\sim
150$~K and decreases sharply below this temperature, indicating that the charged
quasiparticles become decoupled from the collective excitations. Na ordering and
metal-insulator transition for x=0.5 \cite{huang04,zandbergen04} lead to a much more
gradual decrease of $\rm \Gamma(T)$.

\begin{figure}[t]
     \centering
    \includegraphics[height=9cm]{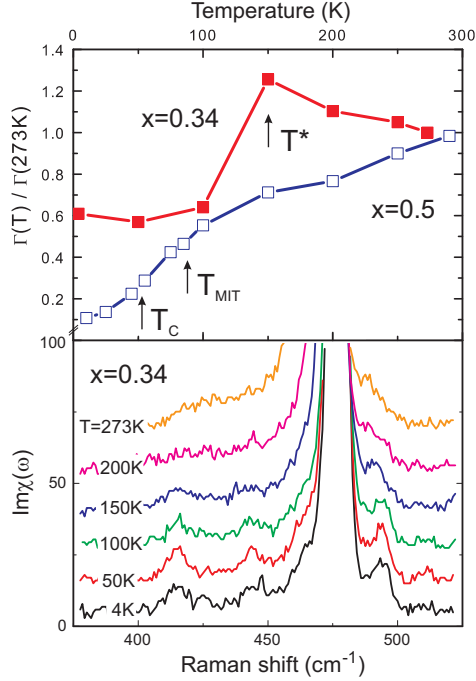}
\caption{Normalized electronic scattering rates $\rm \Gamma(T)$/$\rm \Gamma(273K)$ for
x=0.34/y=1.3 (x=0.5/y=0.0) with full (open) squares with arrows at $T^*$ ($\rm
T_{MIT}$=88K,$\rm T_{c}$=53K) (upper panel) \cite{huang04,zandbergen04}. The lower panel
shows the frequency regime of the $\rm E_{1g}$ phonon with superlattice modes for
T$<$$T^{*}$.
     }
     \label{phonons}
 \end{figure}

The experiments presented here are among the first to address electronic excitations in
SC cobaltates. The reduced intensity of the RS continuum in SC compared to non-SC
samples is in general agreement with prior photoemission experiments indicating a
crossover from strongly to weakly renormalized quasiparticles with decreasing Na
content. Our observation of a pseudogap is also consistent with a recent high-resolution
photoemission experiment that has revealed hints of a partial spectral weight depletion
in $\rm Na_xCoO_2\cdot yH_2O$ in a similar temperature range as that covered by our
experiment \cite{shimojima04}.

A clue to the microscopic origin of the pesudogap is provided by the phonon spectrum in
the lower panel of Fig. 4. Below $T^*$, weak sidebands develop close to the in-plane
$\rm E_{1g}$ phonon, in a manner similar to the charge-ordered state in $\rm
Na_{0.5}CoO_2$ (Fig. 1). In contrast, the out-of-plane mode does not couple strongly to
the electronic states in the plane. In analogy to the latter material, this effect
indicates the formation of a superlattice due to charge ordering. The weak intensity of
these modes in the SC sample and their gradual onset point to a very small order
parameter in the presence of strong fluctuations \cite{additional-modes}. A related
anomaly is observed as an enhanced intensity ratio of the in-plane $\rm E_{1g}$ with
respect to the out-of-plane mode (see lower inset of Fig. 2). This is due to a
renormalization of the phonon intensity by a shielding effect of charge carriers and is
indeed reciprocal to the intensity of the electronic RS itself. Thus, the origin of the
charge ordering instability appears to be predominantly of electronic nature. Evidence
of charge ordering has thus far not been reported in SC samples. However, based on model
calculations charge ordering has been proposed to be an instability competing with SC
\cite{schaak03,baskaran03b,chen04}. In the charge ordered state, the low energy
electronic fluctuations related to the Co $\rm t_{2g}$ states at the Fermi energy are
expected to be partially suppressed, consistent with our observation \cite{koshibae03}.



In conclusion, our observation of a low-energy spectral-weight depletion of the
electronic RS continuum in SC cobaltates provides evidence for the formation of a
pseudogap at temperatures below $T^*$ $\sim 150$~K. Weak phonon anomalies point to the
formation of a charge-ordered state below $T^*$. It is interesting to note that a
pseudogap has recently also been observed in the manganite $\rm
La_{1.2}Sr_{1.8}Mn_{2}O_{7}$ and attributed to polaron dynamics \cite{manella05}. With
respect to the cobaltates, however, further experiments are required to assess whether
charge order coexists microscopically with SC, or whether the material phase-segregates
into SC and charge-ordered states.

We acknowledge discussions with C.~Varma, R.~Hackl, G.~Blumberg, T.~P.~Devereaux,
G.~Khalliulin, C.~Bernhard, R.~Kremer, and Yu.~Pashkevich. This work was supported by
DFG SPP1073, ESF program \emph{Highly Frustrated Magnetism} (ESF-HFM) and the MRSEC
Program of the NSF DMR 02-13282.




\begin{thebibliography}{10}
\bibitem{note} Earlier RS investigations extensively discuss the effects of degradation
\cite{iliev03,lemmens04b} or show a phonon spectrum that deviates from the present
results \cite{shi04}.
\bibitem{additional-modes} The modes at 250, 331, 414, 496~$\mathrm{cm^{-1}}$
show a $\Delta\omega\approx$80~$\mathrm{cm^{-1}}$ comparable to a typical optical phonon
dispersion. The first mode in the proximity to the edge of the depletion has a modified
temperature dependence.
\end{thebibliography}
\end{document}